\documentclass[aps,prl,twocolumn,showpacs,superscriptaddress]{revtex4}
\usepackage{graphicx,epsfig}
\usepackage[usenames]{color}
\usepackage{dsfont,bbm,natbib}
\usepackage{amsfonts}

\usepackage{color}

\begin{document}


\def\beq{\begin{equation}}
\def\eeq{\end{equation}}
\def\be{\begin{equation}}
\def\ee{\end{equation}}

\def\iomn{i\omega_n}
\def\iom#1{i\omega_{#1}} 
\def\c#1#2#3{#1_{#2 #3}}
\def\cdag#1#2#3{#1_{#2 #3}^{+}}
\def\epsk{\epsilon_{{\bf k}}}
\def\Ga{\Gamma_{\alpha}}
\def\Seff{S_{eff}}
\def\dinf{$d\rightarrow\infty\,$}
\def\T{\mbox{Tr}}
\def\t{\mbox{tr}}
\def\cG0{{\cal G}_0}
\def\cS{{\cal S}}
\def\divnum{\frac{1}{N_s}}
\def\vac{|\mbox{vac}\rangle}
\def\intR{\int_{-\infty}^{+\infty}}
\def\intb{\int_{0}^{\beta}}
\def\spinup{\uparrow}
\def\spindown{\downarrow}
\def\bra{\langle}
\def\ket{\rangle}

\def\ka{{\bf k}}
\def\vk{{\bf k}}
\def\vq{{\bf q}}
\def\vQ{{\bf Q}}
\def\vr{{\bf r}}
\def\q{{\bf q}}
\def\R{{\bf R}}
\def\kp{\bbox{k'}}
\def\a{\alpha}
\def\b{\beta}
\def\d{\delta}
\def\D{\Delta}
\def\e{\varepsilon}
\def\eps{\epsilon}
\def\ed{\epsilon_d}
\def\ef{\epsilon_f}
\def\g{\gamma}
\def\G{\Gamma}
\def\l{\lambda}
\def\L{\Lambda}
\def\o{\omega}
\def\ph{\varphi}
\def\s{\sigma}
\def\chib{\overline{\chi}}
\def\et{\widetilde{\epsilon}}
\def\hn{\hat{n}}
\def\hnu{\hat{n}_\uparrow}
\def\hnd{\hat{n}_\downarrow}

\def\hc{\mbox{h.c}}
\def\Im{\mbox{Im}}

\def\est{\varepsilon_F^*}
\def\v2o3{V$_2$O$_3$}
\def\uc2{$U_{c2}$}
\def\uc1{$U_{c1}$}


\def\bea{\begin{eqnarray}}
\def\eea{\end{eqnarray}}
\def \bal{\begin{align}}
\def \eal{\end{align}} 
\def\#{\!\!}
\def\@{\!\!\!\!}

\def\vi{{\bf i}}
\def\vj{{\bf j}}

\def\+{\dagger}


\def\up{\spinup}
\def\down{\spindown}


\def\"{{\prime\prime}}


\title{Orbital Selectivity in Hund's metals: The Iron Chalcogenides}

\author{N. Lanat\`a}
\affiliation{Department of Physics and Astronomy, Rutgers University, Piscataway, New Jersey 08856-8019, USA} 
\author{H. U. R. Strand}
\affiliation{Department of Physics, University of Gothenburg, SE-41296 Gothenburg, Sweden}
\author{G. Giovannetti}
\affiliation{CNR-IOM Democritos National Simulation Center and Scuola Internazionale Superiore di Studi Avanzati (SISSA), Via Bonomea 265, 34136 Trieste, Italy}
\author{B. Hellsing}
\affiliation{Department of Physics, University of Gothenburg, SE-41296 Gothenburg, Sweden}
\author{L. de' Medici}
\affiliation{Laboratoire de Physique des Solides, UMR8502 CNRS-Universit\'{e} Paris-Sud, Orsay, France}
\author{M. Capone}
\affiliation{CNR-IOM Democritos National Simulation Center and Scuola Internazionale Superiore di Studi Avanzati (SISSA), Via Bonomea 265, 34136 Trieste, Italy}

\date{\today}

\begin{abstract}
We show that electron correlations lead to a bad metallic state in 
chalcogenides FeSe and FeTe despite the intermediate
value of the  Hubbard repulsion $U$ and Hund's rule coupling $J$. 
The evolution of the quasi particle weight $Z$ as a function of the 
interaction terms reveals a clear crossover
at $U \simeq$ 2.5 eV. In the weak coupling limit $Z$ decreases for
all correlated $d$ orbitals as a function of $U$ and beyond
the crossover coupling they become weakly dependent on $U$ while 
strongly depend on $J$. A marked orbital
dependence of the $Z$'s emerges even if in general the
orbital-selective Mott transition only occurs for 
relatively large values of $U$. 
This two-stage reduction of the quasi particle coherence due 
to the combined effect of Hubbard $U$ and the Hund's $J$, 
suggests that the iron-based superconductors can be referred to as 
Hund's correlated metals.

\end{abstract}

\maketitle

The role of electron correlations in the iron-based
superconductors is still a debated issue, naturally intertwined
with the search for the origin of high critical temperatures.
We present results that improve the qualitative understanding 
of how electron correlation influences fundamental electron 
properties of these compounds, such as the metallicity, which 
in turn might be important also for the understanding of 
the pairing mechanism. We choose two candidates 
of the chalogenides, FeSe and FeTe and employ 
$first$ $principles$ electron structure calculations combined with 
advanced many-body methods taking into account the 
local electron correlation.
The chalcognides have in contrast to the pnictides a simpler atomic structure, 
thus easier to synthesize and also to study theoretically. In addition they are non toxic in contrast to the pnictides containing arsenic. 

In previously known superconductors we can identify either
weakly correlated materials, like elemental superconductors or binary
alloys, including MgB$_2$, or highly-correlated compound like the
copper oxides and heavy fermion materials. In the first set of
compounds superconductivity is explained within the
Bardeen-Cooper-Schrieffer framework and its extensions, and it occurs
as a pairing instability of a normal metal. In the second set it
is widely believed that correlations revolutionize the electronic
properties and that both the metallic state and the pairing mechanism
deviate from standard paradigms.  

The iron-based pnictides and chalcognides superconductors
do not fit this simple classification. The 
common labeling ``intermediate correlation'',  
referring to properties such as Fermi surface topology or 
absence of Hubbard bands~\cite{Yang_Devereaux_weakcorr}, 
suggests modest effects of correlations. Conversely, the metallic state appears 
much less coherent than what these observations 
imply~\cite{Stewart_RMP,Johnston_Review_FeSC}.
A magnetic counterpart of this dualism is the localized 
an itinerant nature of the spin-density-wave state of 
the parent compound.  

The characteristic property of the band structure is that several of the five $d$-bands cross the Fermi level. The
multi-orbital nature leads to several exotic electronic properties 
such as orbital-selectivity~\cite{demedici_3bandOSMT,demedici_Genesis,Kou_OSMT_pnictides,Hackl_Vojta_OSMT_pnictides,Yin_Weiguo-Spin_fermion,Bascones_OD} 
and also to the conclusion that the
inter-orbital exchange or Hund's coupling plays a key role~\cite{Haule_Hund_pnictides,Georges_annrev}.

The role of the Hund's coupling has indeed been recognized in the
early stages of the field in a Dynamical Mean-Field Theory (DMFT)
study by Haule and Kotliar~\cite{Haule_Hund_pnictides}, 
who coined the definition of Hund's metals by the
observation that the quasi particle effective mass and the
response functions are much more sensitive to the 
Hund's coupling $J$ than to the Hubbard
$U$ interaction. 

For a Hund's metal the spectral weight is not transferred
to the high-energy Hubbard bands, but rather spreads over a scale controlled by $J$.
Other DMFT studies have highlighted the anomalies of the metallic
state, showing its incoherent nature~\cite{Ishida_Mott_d5_nFL_Fe-SC,Liebsch_FeSe_spinfreezing}
and its relation with a spin-freezing crossover~\cite{Werner_122_dynU}.
In Ref.~\cite{Hansmann_localmoment_prl} the dual nature of the
magnetic correlation is shown to induce a remarkable difference between a large
instantaneous magnetic moment and smaller long-time magnetic
correlations, similar to the spin-freezing scenario proposed in
Ref.~\cite{Werner_SpinFreezing} for a three-orbital model.

It has recently been shown that $J$ can have a two-fold effect on a
multiorbital system with an integer filling different from one
electron per orbital~\cite{demedici_Janus}, a situation which is realized in the parent
compounds of iron superconductors, in which six electrons populate the
five $d$ orbitals. In this configuration $J$ reduces the quasi particle
coherence temperature (or coherence energy scale), while it increases
the critical $U$ for the Mott transition. 

As a consequence, a two-stage reduction of the
electronic coherence scale (measured by the quasi particle weight $Z$) occurs
as a function of $U$. Indeed, if we choose a sizable value of $J$ and follow
the evolution of the metallic properties, we
first have a rapid decrease of the effective Fermi-liquid coherence
scale, which leads to a bad metal already for
intermediate correlations strengths, while the Mott transition occurs
only at much larger $U$. This opens a window of $U$ in which $Z$ is
essentially flat, which has been dubbed after the roman god Janus in
view of the double-faced effect of the  Hund's coupling~\cite{demedici_Janus}. 

In this work we explore the combined role of $U$ and $J$ 
in the iron-based chalcogenides FeSe and FeTe by means of the Gutzwiller
approximation (GA). The GA is a simplified treatment of electron correlations 
which systematically 
selects the energetically favorable electronic configurations out of an uncorrelated wave function.
The method provides a reasonable description of the Mott transition from
the metallic side~\cite{brinkman&rice} and allows for a numerically 
cheap investigation of a wide range of model parameters.

We employ the GA numerical scheme developed in Ref.~\cite{Gmethod,DMFTG,full-slater},
which is a generalization of earlier formulations of the GA method~\cite{Michele,Kondo,mybil,Deng_LDA+Gutz} 
and which enables taking into account the full rotationally invariant
Hund's terms, including the so-called spin-flip and pair-hopping, that
are often hard to
treat with approximate analytical methods and even with
numerical methods. Since the formation of a Hund's metal is
actually associated with a differentiation between the different atomic
multiplets, we expect that the GA will perform even better than for
standard Mott transitions.

Based on electronic structure calculations of FeSe and FeTe 
combined with the GA we show that the electronic configuration 
of the parent compounds of iron-based superconductors 
form an ideal system with a two-stage reduction of electronic coherence. 
Furthermore, the bad metal arising from the interplay of $U$
and $J$ displays, as expected, an orbital-selective coherence with
$t_{2g}$ orbitals significantly more correlated than $e_g$.

The material-specific band structure is determined using Density Functional Theory with 
the Generalized Gradient Approximation for the exchange-correlation potential according to the
Perdew-Burke-Ernzerhof recipe as implemented in Quantum
Espresso~\cite{Giannozzi_QE}. Then we apply Wannier90~\cite{Mostofi_Wannier90} to compute
the maximally localized Wannier orbitals, and we include
the interaction terms of the form
\bea\label{H_int}
 H&=&\# U\sum_{i,m}
 n_{im\uparrow} n_{im\downarrow}+(U' -\frac{J}{2})\@\sum_{i,m>m' } n_{im} n_{im'} \\
 &-&\# J\@\sum_{i,m>m'}\@\# \left [ 2 {\bf S}_{im}\#  \cdot {\bf S}_{im'}\!
 +\!(d^\dagger_{im\uparrow}d^\dagger_{im\downarrow}d_{im'\uparrow}d_{im'\downarrow}\#+\text{H.c.})\right]\,.
\nonumber
 \eea
Here $d_{i,m\s}$ is the destruction operator of an electron of spin $\s$ at 
site $i$ in orbital $m$, and $n_{im\s}\equiv d^\+_{im\s}d_{im\s}$, $n_{im}\equiv 
\sum_\s d^\+_{im\s}d_{im\s}$, ${\bf S}_{im}$ is the spin operator for orbital $m$ at site $i$.
$U$ and $U'=U-2J$ are intra- and  inter-orbital repulsions and $J$ is
the Hund's coupling. The values of $U$ and $J$ are not directly accessible from
experiments and even if reliable theoretical estimates are obtained with constrained-RPA,
there are still some discrepancies between different calculations.
In the light of the extreme sensitivity on the value of the parameter $J$, it is particularly useful
to apply a method such as the present GA which allows for a continuous sweep of many parameter
values.


%
%

\begin{figure}
\begin{center}
\includegraphics{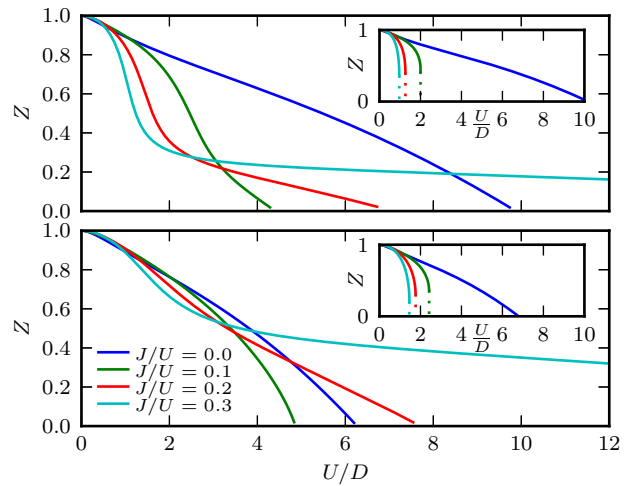}
\caption{Quasiparticle weight in a N$_{\text{orb}}$-fold degenerate Hubbard model
  with average population of N$_{\text{orb}}$+1 electrons, as a function of $U/D$ for
  various Hund's coupling $J/U$, where $D$ is the half-bandwidth. Upper panel: N$_{\text{orb}}$=5, Lower panel
  N$_{\text{orb}}$=3. The panels show data for N=N$_{\text{orb}}$+1 electrons, the
  insets for N=N$_{\text{orb}}$.}\label{fig:deg_model}
\end{center}
\end{figure}

As mentioned above, one of the distinctive features of the iron-based
superconductors is that all five $d$-orbitals appear to contribute to the 
band structure around the Fermi level. As a first step we 
consider a system with five degenerate $d$ bands, and show that the
configuration with six electrons per atom, characteristic of the parent
compounds, is a clear cut case of a ``Janus'' scenario, characterized by a two-stage
reduction of the quasi particle weight.

In Refs. \cite{demedici_Janus,Georges_annrev} it is clearly shown that the two-stage reduction of the
electronic coherence scale is a consequence of a contrasting effect of
$J$ on the metallic character of the electrons. In the weak-coupling limit 
$J$ favors the formation of a large local magnetic moment, which leads to
a faster decay of the electronic coherence scale $Z$, while in the
strong-coupling the Mott transition is pushed to larger $U$. 
This effect is particularly strong when the number of electrons per
atom differs by one unit from the number of orbitals 
$N = N_{\text{orb}} \pm 1$, and it is expected to be emphasized
increasing the number of orbitals as the weak-coupling coherence temperature scales
exponentially with $N_{\text{orb}}$. 

In Fig.~\ref{fig:deg_model} we compare the GA results 
for the two cases of $N = N_{\text{orb}} \pm 1$ when $N_{\text{orb}} = 5$ and $N_{\text{orb}} = 3$. 
We clearly see that the former case has a much
clearer separation between a regime in which $Z$ rapidly decreases as
a function of $U$ and a large bad metal region in which $Z$ is essentially constant 
prior to the Mott insulator transition. In the inset we show the case of 
half-filling, $N = N_{\text{orb}}$, where no dual nature is observed.

Once established that the electron count of the parent
compounds of the iron-based superconductors gives rise to a strongly two-faced
correlation physics, we move towards the realistic situation in order
to identify how the material-specific properties influence this
picture.
\begin{figure}
\begin{center}
\includegraphics{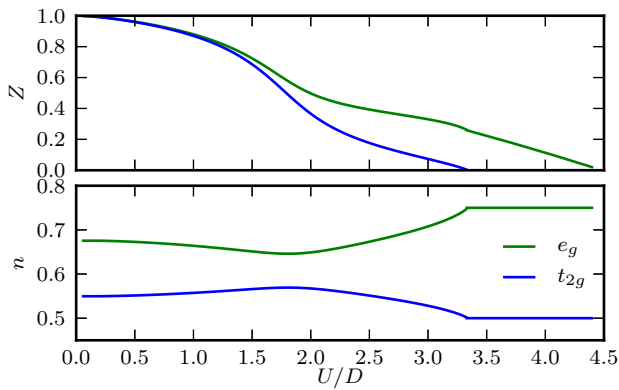}
\caption{Upper panel: quasiparticle weight in a Hubbard model with 6 electrons in 5 
bands with semi-circular densities with half-bandwidth $D$   
split by a cubic crystal-field in two manifolds of degeneracy 3 ($t_{2g}$ symmetry) and 2 
($e_g$ symmetry). Lower panel: populations of the two manifolds.}\label{fig_cf}
\end{center}
\end{figure}
As an intermediate step, we lift the degeneracy with a cubic crystal-field which is present in iron pnictides and chalcogenides.
An energy splitting $\Delta$ is introduced between the three $t_{2g}$ and the two $e_g$ orbitals. In Fig. \ref{fig_cf} we
show the results for $\Delta/D = 0.2$. We observe
that, while the weak-coupling region gives an essentially
orbital-independent $Z$, as soon as we enter in the strongly
correlated region, the low-lying states become more correlated than
the higher-lying. In other words, the crystal-field triggers a strongly
orbital-selective renormalization in the bad metal state.

\begin{figure}
\begin{center}
\includegraphics{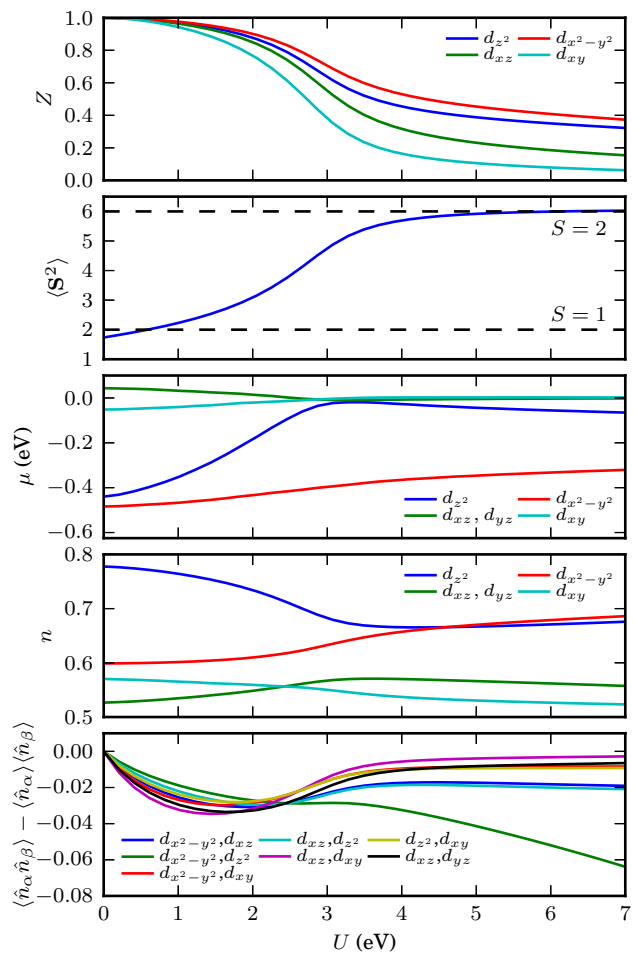}
\caption{Results for FeSe  (DFT+GA with $J/U=0.224$) as a function of $U$. From the top; panel (a), 
the quasi particle weights for the different orbitals, in panel 
(b) the expectation value of $S^2$, in panel (c) the renormalized 
crystal-field splittings, in panel (d) the population of each 
orbital, in panel (e) the inter-orbital density correlations.}\label{fig_FeSe}
\end{center}
\end{figure}

We finally perform the realistic DFT+GA calculation for iron
chalcogenides. In panel (a) of Fig.~\ref{fig_FeSe} we show the evolution of the quasi particle
weight for the different orbitals as a function of $U$, keeping the
ratio $J/U$ fixed to 0.224. This ratio is chosen according to the 
estimates presented in Ref.~\cite{Aichhorn_FeSe} for FeSe.
The picture remains similar to the idealized systems. For small values of $U$ the $Z$'s for the 
different orbitals are similar and they appreciably decrease as a
function of $U$ before $U \simeq 2.5$ eV, where the system enters the
novel regime in which the $Z$'s are small and almost constant as a function of
$U$. However, an orbital dependence also appears clearly. In addition to the differentiation of the $t_{2g}$ and
$e_g$ orbitals, we find that the $d_{xy}$ orbital is the most
correlated and the $d_{x^2-y^2}$ is more localized than the $d_{3z^2-r^2}$.
The crossover, which roughly separates a weakly-correlated phase from
a bad metallic phase, takes place at a value of $U$ smaller than the bandwidth 
$2D$ ($\sim$ 4eV), and much
smaller than the multiband Mott transition, which would take place 
at a $U$ of the order of 5 times the width of each band~\cite{demedici_MottHund}.

It is easy to see that in the atomic limit the ground state changes from low-spin
($S=1$) to high-spin ($S=2$) when $J$ becomes larger than the
crystal-field splittings ($\sim0.6eV$ for FeSe). In the metallic phase
this evolution yields a crossover to a $S=2$ state (Panel (b)) which leads to the rapid reduction of
$Z$~\cite{Georges_annrev}.

This crossover leads to a dramatic lowering of the coherence temperature,
and opens a wide bad-metal region due to the increase of
$U_c$ induced by the effect of $J$ on the high-spin Mott gap. This behavior is
observed even more pronounced in studies of
LaFeAsO~\cite{YuSi_LDA-SlaveSpins_LaFeAsO} and the intercalated
chalcogenides~\cite{Yu_Si_KFeSe}. It contributes in a substantial
manner to the sharp onset of the Hund's
metal~\cite{Haule_Hund_pnictides}/spin-frozen~\cite{Werner_122_dynU}/incoherent~\cite{Ishida_Mott_d5_nFL_Fe-SC,Liebsch_FeSe_spinfreezing}
phase
 observed in all DMFT studies.

This effect is also reflected in the renormalized orbital
energies (panel (c) of Fig.~\ref{fig_FeSe}), with four of the five
orbitals being brought close to one another and also near the Femi level by the interactions. 
This favors a more even population of
the orbitals that gains in exchange
energy, and favors the high-spin configurations over the low-spin ones.
The evolution of the population of the different orbitals is shown in 
panel (d) of Fig.~\ref{fig_FeSe}.
The $t_{2g}$ orbitals have populations closer to
half-filling already at the DFT level, while the $e_g$ bands are more
occupied. Increasing the interaction the $d_{xy}$ level, which has the
smaller $Z$, becomes less occupied than $d_{xz}$ and $d_{yz}$ due
to the stronger effect of correlations. Analogously, the large
difference between the non-interacting densities of the two $e_g$
bands is washed out by correlations, which favor a more democratic
occupation with high spin.  The same low to high spin transition within the metallic phase occurs
for any sizable value of $J$ and it is indeed present already in the model with a simple $t_{2g}$-$e_g$ 
splitting as clear from the non-monotonic population behavior shown in the lower panel of Fig.~\ref{fig_cf}. 

In Panel (e) of Fig.~\ref{fig_FeSe}  
we show the inter orbital density correlation functions, which are
clearly suppressed in the correlated regime. This suppression, driven by $J$, has been put
forth~\cite{demedici_3bandOSMT, demedici_MottHund} as the driving
mechanism behind the orbital selectivity. Indeed $J$ acts as an
``orbital decoupler'' (``band decoupler''~\cite{demedici_MottHund},
for weak orbital hybridization) suppressing inter-orbital charge fluctuations, 
and rendering the charge dynamics of each orbital virtually independent.
\begin{figure}
\begin{center}
\includegraphics{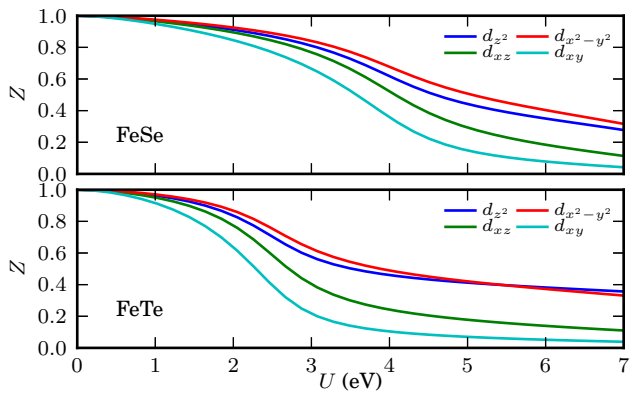}
\caption{Results for FeSe and smaller $J/U=0.15$ (top panel) and for
  FeTe (bottom panel)}\label{fig_FeTe}
\end{center}
\end{figure}
For a smaller value of $J/U$, 0.15, the picture does not change. 
The position of the crossover is only weakly affected, while 
the values of the $Z$'s in the bad metallic region after the crossover
depend strongly on $J$. The behavior observed confirms previous
findings of a $Z$ which depends strongly on $J$ and weakly on $U$ in
the physically relevant region of $U \sim 4$ eV and $J \sim 0.5-1$
eV. At the same time our results clearly underline that such ``Hund's
metal'' requires a critical value of the Hubbard repulsion, albeit
much smaller than one might expect on the basis of the value of the
bandwidth. The picture is clearly consistent with that drawn in
Ref.~\cite{demedici_Janus}. Finally, we show the quasi particle weights for FeTe using the same
value of the interaction coefficients. The main difference is a sharper
separation between $t_{2g}$ and $e_g$ orbitals, and a larger
renormalization for the $e_g$ orbitals. 

In summary, we have calculated the correlation strength induced by
many-body correlations on the ab-initio electronic structure of the
iron chalcogenides FeSe and FeTe. We find, in agreement with previous analogous studies on 
LaFeAsO~\cite{YuSi_LDA-SlaveSpins_LaFeAsO} and K$_{1-x}$Fe$_{2-y}$Se$_2$~\cite{Yu_Si_KFeSe}, that 
Hund's coupling has a strong influence on the electronic properties of the paramagnetic phase, 
inducing a two-stage quasi particle renormalization. A first regime at weak coupling sees a moderate 
correlation affecting all orbitals comparably. After a quick decrease around $U\simeq 2.5eV$, a value 
much smaller than the overall bandwidth, a strongly correlated regime is entered, heavily differentiated 
among the orbitals (with $t_{2g}$ orbitals sensibly more correlated), in which the quasi particle weights are almost independent of $U$. 
The Mott transition occurs at much higher ($\sim5$ times the bandwidth) interaction strengths.
Comparison with idealized models shows that the
two-staged reduction of the quasi particle weights is due to the
filling of 6 electrons in 5 bands, thus placing the system in the
``Janus'' regime induced by Hund's coupling~\cite{demedici_Janus}, and
that, by introducing a crystal-field $t_{2g}$-$e_g$ splitting, orbital differentiation happens once entered the Janus
regime, where the orbitals closest to half-filling are more correlated~\cite{demedici_MottHund}.

We acknowledge useful discussions with L. Bascones. 
M.C. and G.G. acknowledge financial support of FP7/ERC through Starting Independent Research Grant ``SUPERBAD'' (Grant Agreement n. 240524).
H.U.R.S acknowledges funding from the Mathematics Physics Platform (MP2) at the University of Gothenburg. 
L.d.M. acknowledges funding from Agence Nationale de la Recherche (project ANR-09-RPDOC-019-01).
The calculations were partly performed on resources provided by the Swedish National Infrastructure for Computing (SNIC) at
Chalmers Centre for Computational Science and Engineering
(C3SE) (project 001-10-37).

\bibliographystyle{apsrev}

\end{document}